\documentclass[hidelinks,conference]{ieeeconf}      

\IEEEoverridecommandlockouts                              

\usepackage{xr}

\newcommand{\omitted}[1]{}

\title{
 \fontsize{22}{22} \selectfont 
No-Regret Model Predictive Control \\
with Online Learning of Koopman Operators

}

\author{
Hongyu Zhou \;\; Vasileios Tzoumas
\thanks{This work was partially supported by NSF CAREER No. 2337412.}
\thanks{Department of Aerospace Engineering, University of Michigan, Ann Arbor, MI 48109 USA;  {\tt\footnotesize \{zhouhy, vtzoumas\}@umich.edu}}
\vspace{6mm}
}

\usepackage{cite}

\usepackage{comment}
\usepackage{siunitx}
\usepackage{relsize}
\usepackage{ifthen}
\usepackage[colorinlistoftodos]{todonotes}
\usepackage[vlined,ruled,linesnumbered]{algorithm2e}
\usepackage{graphics} 
\usepackage{rotating}
\usepackage{color}
\usepackage{enumerate}
\usepackage[T1]{fontenc}
\usepackage{psfrag}
\usepackage{epsfig} 
\usepackage{booktabs}
\usepackage{graphicx}
\usepackage{url}
\usepackage{multirow}
\usepackage{array}
\usepackage{latexsym}
\usepackage{amsfonts}
\usepackage{amsmath}
\usepackage{amssymb}
\usepackage{xstring}
\usepackage{algorithmic}
\usepackage{multirow}
\usepackage{xcolor}
\usepackage{prettyref}
\usepackage{flexisym}
\usepackage{bigdelim}
\usepackage{breqn} 
\usepackage{listings}
\let\labelindent\relax
\usepackage{xspace}
\usepackage{bm}
\graphicspath{{./figures/}}
\usepackage{tikz}
\usetikzlibrary{matrix,calc}
\usepackage{lipsum}
\usepackage{mdwlist}

\let\itemize\undefined
\makecompactlist{itemize}{stditemize}
\usepackage{enumitem}
\usepackage{caption}
\usepackage{epstopdf}
\usepackage{subfigure}

\usepackage{amsthm}
\usepackage{mathtools}

\makeatletter
\let\NAT@parse\undefined
\makeatother
\usepackage{hyperref}
\usepackage{cleveref}

\hypersetup{%
  colorlinks=true,
  linkcolor=blue,
  filecolor=magenta,      
  urlcolor=black,
  citecolor=red,
  linkbordercolor={0 0 1}
}

\newtheorem{theorem}{Theorem}
\newtheorem{problem}{Problem}

\newtheorem{assumption}{Assumption}
\newtheorem{definition}{Definition}
\newtheorem{proposition}{Proposition}

\newtheorem{remark}{Remark}

\newcommand{\bdmath}{\begin{dmath}}
\newcommand{\edmath}{\end{dmath}}
\newcommand{\beq}{\begin{equation}}
\newcommand{\eeq}{\end{equation}}
\newcommand{\bdm}{\begin{displaymath}}
\newcommand{\edm}{\end{displaymath}}
\newcommand{\bea}{\begin{eqnarray}}
\newcommand{\eea}{\end{eqnarray}}
\newcommand{\beal}{\beq \begin{array}{lll}}
\newcommand{\eeal}{\end{array} \eeq}
\newcommand{\beas}{\begin{eqnarray*}}
\newcommand{\eeas}{\end{eqnarray*}}
\newcommand{\ba}{\begin{array}}
\newcommand{\ea}{\end{array}}
\newcommand{\bit}{\begin{itemize}}
\newcommand{\eit}{\end{itemize}}
\newcommand{\ben}{\begin{enumerate}}
\newcommand{\een}{\end{enumerate}}

\newcommand{\calD}{{\cal D}}

\newcommand{\calF}{{\cal F}}

\newcommand{\calO}{{\cal O}}

\newcommand{\calU}{{\cal U}}

\definecolor{myblue}{RGB}{65 105 225}

\newcommand{\hide}[1]{}

\newcommand{\hiddenText}{{\color{gray} hidden text.}}
\newcommand{\hideWithText}[1]{\hiddenText}

\newcommand{\diag}[1]{\mathrm{diag}\left(#1\right)}

\newcommand{\scenario}[1]{{\fontsize{9}{8.7}\selectfont\sf#1}\xspace}
\newcommand{\scenariot}[1]{{\fontsize{8}{8}\selectfont\sf#1}\xspace}

\newcommand{\myx}{\mathbf{x}}

\newcommand{\ie}{\emph{i.e.},\xspace}
\newcommand{\eg}{\emph{e.g.},\xspace}

\newcommand{\myParagraph}[1]{{\bf #1.}\xspace}

\newcommand{\OCO}{\scenario{{OCO}}}

\newcommand{\OGD}{\scenario{{OGD}}}
\newcommand{\GP}{\scenario{{GP}}}

\newcommand{\MPC}{\scenario{{MPC}}}
\newcommand{\NMPC}{\scenario{{Nominal MPC}}}
\newcommand{\GPMPC}{\scenario{{GP-MPC}}}

\newcommand{\KMMPC}{\scenario{{Koopman-MPC}}}
\newcommand{\RFFMPC}{\scenario{{RFF-MPC}}}

\newcommand{\SReg}{\operatorname{Regret}_T^S}
\newcommand{\DReg}{\operatorname{Regret}_T^D}

\PassOptionsToPackage{end}{algorithmic}

\begin{document}

\maketitle

\thispagestyle{empty}
\pagestyle{empty}

\begin{abstract}
We study a problem of simultaneous system identification and model predictive control of nonlinear systems. 
Particularly, we provide an algorithm for systems with unknown residual dynamics that can be expressed by Koopman operators.  
Such residual dynamics can model external disturbances and modeling errors, such as wind and wave disturbances to aerial and marine vehicles, or inaccurate model parameters.
The algorithm has finite-time near-optimality guarantees and asymptotically converges to the optimal non-causal controller.  
Specifically, the algorithm enjoys sublinear \textit{dynamic regret}, defined herein as the suboptimality against an optimal clairvoyant controller that knows how the unknown  dynamics will adapt to its states and actions.
To this end, we assume the algorithm is given Koopman observable functions such that the unknown dynamics can be approximated by a linear dynamical system. 
Then, it employs model predictive control based on the current learned model of the unknown residual dynamics.
This model is updated online using least squares in a self-supervised manner based on the data collected while controlling the system.
We validate our algorithm in physics-based simulations of a cart-pole system aiming to maintain the pole upright despite inaccurate model parameters.
\end{abstract}


\section{Introduction}\label{sec:Intro}

In the future, mobile robots will leverage their on-board
control capabilities to complete tasks such as package delivery~\cite{ackerman2013amazon},  target tracking~\cite{chen2016tracking}, and inspection and maintenance~\cite{seneviratne2018smart}.
Such tasks require accurate and efficient tracking control under uncertainty, particularly, under unknown dynamics and external disturbances. This is challenging since the uncertainty is versatile across different environments and is possibly adaptive to robots' actions and states.  
Examples of such tasks are: quadrotors to (i) pick up and carry packages of unknown weight, (ii) chase a moving target at high speeds where the induced aerodynamic drag is hard to model, and (iii) inspect and maintain outdoor infrastructure exposed to turbulence and wind gusts.

\begin{figure}[t]
    \centering
    \includegraphics[width=0.49\textwidth]{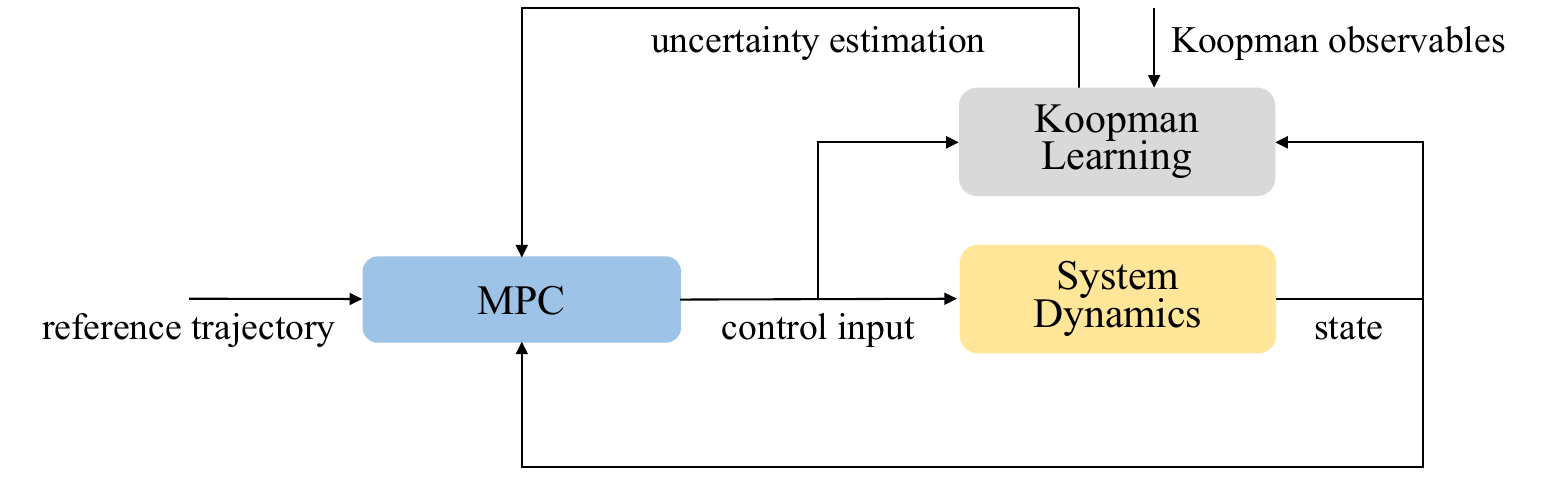}
    \captionsetup{font=small}
    \caption{\textbf{Overview of Pipeline for Model Predictive Control with Online Learning of Koopman Operator.} The pipeline is composed of two interacting modules: (i) a model predictive control (\scenariot{MPC}) module, and (ii) an online Koopman learning module with predefined Koopman observable functions. The \scenariot{MPC} module uses the estimated unknown dynamics from the Koopman learning module to calculate the next control input. Given the control input and the observed new state, the online Koopman learning module then updates the estimate of the unknown dynamics. }
    \label{fig_framework}
\end{figure}

State-of-the-art methods for control under unknown dynamics and disturbances include:   
robust control~\cite{mayne2005robust,goel2020regret,martin2024guarantees,didier2022system,zhou2023safe}; 
adaptive control and disturbance compensation~\cite{slotine1991applied,krstic1995nonlinear,tal2020accurate,wu2023mathcal,das2024robust}; 
and online learning~\cite{hazan2022introduction,zhou2023safecdc,zhou2023saferal,zhou2023efficient,boffi2021regret,nonhoff2024online,tsiamis2024predictive,zhou2025simultaneous,zhou2025adaptive}. 
The robust control methods, given a known upper bound on the magnitude of the noise, can be conservative due to assuming worse-case disturbance realization~\cite{zhou1998essentials}, instead of planning based on an accurate predictive model of the disturbance. 
Similarly, the adaptive control and the online learning control methods may exhibit sub-optimal performance due to only reacting to the history of observed disturbances, instead of planning based on an accurate predictive model of the disturbance~\cite{slotine1991applied,krstic1995nonlinear,hazan2022introduction}.
Relevant work on simultaneous system identification and model predictive control are~\cite{zhou2025simultaneous,zhou2025adaptive}, where the model of unknown dynamics and disturbances are learned by random Fourier features~\cite{rahimi2007random}. 
The method is suitable when we have no prior knowledge about the unknown dynamics, \ie no knowledge about what features are representative to approximating the unknown dynamics, since random Fourier features can be sampled from predefined distributions. 
In the case where we can identify suitable features, control performance can be benefited by using such features to learn unknown dynamics.


In this paper, we leverage the success of Koopman operator in modeling and learning of nonlinear dynamics~\cite{koopman1931hamiltonian}. 
Koopman operator represents nonlinear dynamics as a (potentially infinite-dimensional) linear system by evolving functions of the state of interest, \ie Koopman observables, in time. Such representations can be incrementally updated computationally efficiently, enabling online learning of Koopman operator. Moreover, due to the linear representation, the learned model can be incorporated into model predictive control for real-time applications.
Therefore, we propose a self-supervised method to learn online a predictive model of the unknown uncertainties approximated by Koopman operator~(Fig.~\ref{fig_framework}). The proposed method promises to enable: one-shot online learning (as opposed to offline or episodic learning); online adaptation to the actual disturbance realization (as opposed to the worst-case); and control planned over a look-ahead horizon of predicted system dynamics and disturbances (as opposed to the their past).  We elaborate on our contributions next.

\myParagraph{Contributions} 
We provide a real-time and asymptotically optimal algorithm for the simultaneous Koopman learning and control of nonlinear systems. 
The algorithm is self-supervised and learns unknown residual dynamics that can be expressed by Koopman operator.  
Specifically, the algorithm is composed of two interacting modules (Fig.~\ref{fig_framework}): (i) a Model Predictive Control (\MPC) module, and (ii) an online Koopman learning module with pre-specified Koopman observable functions.  At each time step, the \MPC module uses the estimated unknown residual dynamics from the Koopman learning module to calculate the next control input. Given the control input and the observed new state, the online Koopman learning module then updates the estimate of the unknown dynamics by least-squares estimation via online gradient descent (\OGD)~\cite{hazan2016introduction}.  


The algorithm has \textit{no dynamic regret}, that is,  it asymptotically matches the performance of the optimal controller that knows a priori the unknown dynamics.\footnote{This definition of dynamic regret differs from the standard one in the non-stochastic control literature, \eg \cite{hazan2022introduction,zhou2023efficient,zhou2023safecdc,zhou2023saferal},  where the optimal controller and the Algorithm experienced the same realization of disturbances, instead of different realizations given that the unknown dynamics can be adaptive to states and actions~(\Cref{rem:adaptivity}).} Particularly, we provide the following finite-time performance guarantee for \Cref{alg:MPC} (\Cref{theorem:regret_OLMPC}):
    $$\text{Regret of \MPC} \leq \calO\left(T^{\frac{3}{4}}\right).$$



\myParagraph{Numerical Evaluations}
We validate the algorithm in physics-based simulations~ in simulated scenarios of a cart-pole system that aims to stabilize at a setpoint despite inaccurate model parameters (\Cref{sec:exp-sim}).
We compare our algorithm with a nominal \MPC (\NMPC) that ignores the unknown dynamics or disturbances, the Gaussian process \MPC (\GPMPC)~\cite{hewing2019cautious}, and the \MPC with random Fourier features (\RFFMPC)~\cite{zhou2025simultaneous}.
Our method achieves better tracking performance than \GPMPC and \RFFMPC. 
\section{Related Work}\label{sec:lit_review}

We discuss work on adaptive control; robust control; non-stochastic control; and Koopman operator for control.

\paragraph*{Adaptive control}
Adaptive control methods often assume parametric uncertainty additive to the known system dynamics, \eg parametric uncertainty in the form of unknown coefficients multiplying known basis functions~\cite{slotine1991applied,krstic1995nonlinear}, or directly estimate the value of unknown disturbances~\cite{tal2020accurate,wu2023mathcal,das2024robust}. They update online the coefficients or the value of unknown disturbances, and generate an adaptive control input to compensate for the estimated disturbances. 
In contrast, we leverage the Koopman operator to learn a model of the residual dynamics online to enable adaptive model predictive control.

\paragraph*{Robust control}
Robust control algorithms select control inputs based on the assumption of worst-case realization of disturbances with an upper bound of its magnitude~\cite{mayne2005robust,goel2020regret,martin2024guarantees,didier2022system,zhou2023safe}. 
However, assuming the worst-case disturbances can be conservative. 
In this paper, instead, we incorporate a learned model of disturbances to reduce conservativeness caused by the worst-case disturbances assumption.


\paragraph*{Non-stochastic control} 
Online learning algorithms, also known as non-stochastic control algorithms, provide controllers with bounded regret guarantees, upon employing the \OCO framework to capture the control problem as a sequential game between a controller and an adversary~\cite{hazan2022introduction,boffi2021regret,zhou2023safecdc,zhou2023saferal,zhou2023efficient,nonhoff2024online,tsiamis2024predictive}. 
They typically select control inputs based on past information~\cite{hazan2022introduction,boffi2021regret,zhou2023safecdc,zhou2023saferal,zhou2023efficient,nonhoff2024online}, instead of using \MPC as in this paper.
\cite{tsiamis2024predictive} provides a \MPC method for target tracking, focusing on learning unknown trajectories of the target, as opposed to learning unknown dynamics/disturbances in our paper. Additionally, it focuses on linear systems and linear \MPC.  Instead, we utilize nonlinear \MPC for control-affine systems.
\cite{zhou2025simultaneous,zhou2025adaptive} learn the residual dynamics with random Fourier features~\cite{rahimi2007random} for \MPC, where those features are sampled randomly from prespecified distributions. The method is suitable when we have no prior knowledge about the unknown dynamics and what features are representative to approximating the unknown dynamics. 
By contrast, we leverage Koopman operator theory to learn residual dynamics with properly selected Koopman observable functions, and demonstrate better performance than \cite{zhou2025simultaneous}.

\paragraph*{Koopman Operator for control}
Koopman operator~\cite{koopman1931hamiltonian} represents a nonlinear dynamical system as a (potentially infinite-dimensional) linear system by evolving functions of the state, namely, observables, in time. 
Such Koopman dynamics can be found through
data-driven methods~\cite{brunton2021modern,shi2024koopman,kaiser2020data} that generate a finite-dimensional approximation to the theoretical infinite-dimensional Koopman operator. Their methods typically aim to represent the whole system dynamics by Koopman operator learned offline~\cite{brunton2016sparse,proctor2016dynamic,proctor2018generalizing,korda2018linear,bruder2020data,abraham2019active,folkestad2021koopman}, while we focus only on the online learning of residual dynamics. 
We can enable online adaptation in the case of distribution shift between the data collected offline and encountered during employment, upon combining with those methods of offline Koopman operator learning~(\Cref{remark:offline}). 
We also provide a no-dynamic-regret guarantee for the provided simultaneous Koopman learning and \MPC algorithm.

\section{Model Predictive Control with \\ Online Learning of Koopman Operator}\label{sec:problem}

We formulate the problem of \textit{Model Predictive Control with Online Learning of Koopman Operator}~(\Cref{prob:control}).
To this end, we use the following framework and assumptions.

\myParagraph{Control-Affine Dynamics}
We consider control-affine system dynamics of the form
\begin{equation}
    x_{t+1} = f\left(x_{t}\right) + g\left(x_{t}\right) u_{t} + w_{t}, \quad t \geq 1, 
    \label{eq:affine_sys}
\end{equation}
where $x_t \in\mathbb{R}^{d_x}$ is the state, $u_t \in\mathbb{R}^{d_u}$ is the control input, $f: \mathbb{R}^{d_x} {\rightarrow} \mathbb{R}^{d_x}$, $g: \mathbb{R}^{d_x} {\rightarrow} \mathbb{R}^{d_x} \times \mathbb{R}^{d_u}$ are known locally Lipschitz functions, $w_{t} \triangleq h\left(w_{t-1},z_{t}\right): \mathbb{R}^{d_x}\times\mathbb{R}^{d_z} \rightarrow \mathbb{R}^{d_x}$ is an unknown locally Lipschitz function {with bounded magnitude}, and $z_t \in\mathbb{R}^{d_z}$ is a vector of features chosen as a subset of $[x_t^\top \ u_t^\top]^\top$.

We refer to the undisturbed $x_{t+1} = f\left(x_{t}\right) + g\left(x_{t}\right) u_{t}$ as the \textit{nominal dynamics}, and $w_{t}$ or $h$ as the \textit{residual dynamics}. $w_{t}$ represents unknown disturbances or system dynamics. 
Examples of such unknown disturbances or system dynamics are given in~\cite{jia2023evolver}.


\myParagraph{Model Predictive Control (\MPC)} 
\MPC selects a control input $u_t$ by simulating the nominal system dynamics over a look-ahead horizon $N$, \ie~\MPC selects $u_t$ by solving the optimization problem~\cite{rawlings2017model}: 
\begin{subequations}
    \label{eq:mpc_def}
    \begin{align}
        &\hspace{-4mm}\underset{{u}_{t}, \ \ldots, \ {u}_{t+N-1}}{\textit{min}} \hspace{4mm}\sum_{k=t}^{t+N-1} c_{k}\left(x_{k},u_{k}\right) \label{eq:mpc_def_obj} \\
        & \ \ \operatorname{\textit{subject~to}} \;\quad x_{k+1} = f\left(x_{k}\right) + g\left(x_{k}\right) u_{k},\\
        & \qquad \qquad \qquad \; u_{k}\in \calU, \ \ k\in\{t,\ldots, t+N-1\},
    \end{align}
\end{subequations}
where $c_{t}\left(\cdot,\cdot\right): \mathbb{R}^{d_{x}} \times \mathbb{R}^{d_{u}} {\rightarrow} \mathbb{R}$ is the cost function, and $\calU$ is a compact set that represents constraints on the control input due to, \eg controller saturation.

The optimization problem in \cref{eq:mpc_def} ignores the residual dynamics $h$. To improve performance in the presence of $h$, in this paper we propose a method to estimate $h$ online so \cref{eq:mpc_def} can be adapted to the optimization problem:
\begin{subequations}
    \label{eq:mpc_ada_def}
    \begin{align}
       &\hspace{-4mm}\underset{{u}_{t}, \ \ldots, \ {u}_{t+N-1}}{\textit{min}} \hspace{4mm} \sum_{k=t}^{t+N-1} c_{k}\left(x_{k},u_{k}\right) \label{eq:mpc_ada_obj} \\
        & \ \ \operatorname{\textit{subject~to}} \;\quad x_{k+1} = f\left(x_{k}\right) + g\left(x_{k}\right) u_{k} + \hat{w}_{t}, \label{eq:mpc_ada_dyn}\\
        & \qquad \qquad \qquad \ u_{k}\in \calU,  \ \ k\in\{t,\ldots, t+N-1\}, 
    \end{align}
\end{subequations}
where $\hat{w}_{t}$ is the estimate of ${w}_{t}$.  Specifically, $\hat{w}_{t} \triangleq {h}\left(\cdot,~\cdot~; \hat{\alpha}\right)$ where $\hat{\alpha}$ is the parameter that is updated online by our proposed method to improve the control performance.

We define the notion of \MPC's \textit{value function} and state the assumption on the cost function and value function.

\begin{definition}[Value Function~\cite{grimm2005model}]\label{def:opt_value}
Given a state $x$ and parameter $\hat{\alpha}$, the \emph{value function} $V_t\left(x ; \hat{\alpha}\right)$  is defined as the optimal value of \cref{eq:ada_mpc_value_obj_mother}:
\begin{subequations}
\label{eq:ada_mpc_value_obj_mother}
    \begin{align}
        &\hspace{-4mm}\underset{{u}_{t}, \ \ldots, \ {u}_{t+N-1}}{\textit{min}} \hspace{4mm} \sum_{k=t}^{t+N-1} c_{k}\left(x_{k},u_{k}\right) \label{eq:ada_mpc_value_obj} \\
        & \ \ \operatorname{\textit{subject~to}} \;\quad x_{k+1} = f\left(x_{k}\right) + g\left(x_{k}\right) u_{k} + \hat{w}_{t}, \label{eq:ada_mpc_value_dyn}\\
        & \qquad \qquad \qquad \ x_{t} = x, \; u_{k}\in \calU,  \;  k\in\{t,\ldots, t+N-1\}.
    \end{align}
    \label{eq:ada_mpc_value}
\end{subequations}
\end{definition}

\begin{assumption}[Bounds on Cost Function and Value Function~\cite{grimm2005model}]\label{assumption:stability}
    There exist positive scalars $\underline{\lambda}$, $\overline{\lambda}$, and a continuous function $\sigma: \mathbb{R}^{d_x} \rightarrow \mathbb{R}_{+}$, such that
    (i) $c_{t}\left(x,u\right) \geq \underline{\lambda} \sigma\left(x\right)$, $\forall x,\; u,\; t$; 
    (ii) $V_t\left(x; \hat{\alpha}\right) \leq \overline{\lambda}\sigma\left(x\right)$, $\forall x,\; t$,
    and (iii) $\lim_{\|x\| \rightarrow \infty} \sigma\left(x\right) \rightarrow \infty$.
\end{assumption} 

Under \Cref{assumption:stability}, the \MPC policy in \cref{eq:mpc_ada_def} can be proved to ensure that the system in \cref{eq:mpc_ada_dyn} is globally asymptotic stable \cite{grimm2005model}. 

A cost function that satisfies \Cref{assumption:stability} is the quadratic cost $c_{t}\left(x_{t}, u_{t}\right) = x_{t} Q x_{t}^\top + u_{t} R u_{t}^\top$ when, for example, the system dynamics is linear~\cite[Lemma~1]{grimm2005model}, 
or when the quadratic cost is (exponentially/asymptotically) controllable to zero with respect to $\sigma: \mathbb{R}^{d_x} \rightarrow \mathbb{R}_{+}$~\cite[Section. III]{grimm2005model}.

\begin{assumption}[Lipschitzness]\label{assumption:lipschitz}
    We Assume that ${c}_t\left(x, u\right)$ is locally Lipschitz in $x$ and $u$, $\hat{h}\left(\cdot\right)$ is locally Lipschitz in $\hat{\alpha}$.
\end{assumption} 
\Cref{assumption:lipschitz} will be used to establish the Lipschitzness of the value function $V_t\left(x ; \hat{\alpha}\right)$ with respect to the initial state $x$ and parameter $\hat{\alpha}$.

\myParagraph{Koopman operator}
A Koopman operator~\cite{koopman1931hamiltonian} represents a nonlinear dynamical system $h$ by a (potentially infinite-dimensional) linear system.
Consider first the residual dynamics without dependence on $z_t$: $w_{t} \triangleq h\left(w_{t-1}\right)$; and define the observable function $\Phi \in \calO$, where $\calO$ is the infinite-dimensional function space of all observation functions. Then, the Koopman operator $\kappa: \calO \rightarrow \calO$ is an operator acting on $\Phi$ such that
\begin{equation}
     \Phi\left( w_{t+1} \right) = \kappa \Phi\left( w_{t} \right).
\end{equation}

Typically, $\kappa$ cannot be implemented due to infinite dimensionality. Instead, a finite subspace approximation $A_k \in \mathbb{R}^{d_\Phi} \times \mathbb{R}^{d_\Phi}$ acting on a subspace $\calF \subset \calO$ is used such that 
\begin{equation}
     \Phi\left( w_{t+1} \right) = A_k \Phi\left( w_{t} \right) + r, 
     \label{eq:koopman_auto}
\end{equation}
where $r \in \calO$ is the error due to a finite dimensional approximation of $\kappa$. In principle, the error $r \rightarrow 0$ as $\calF \rightarrow \calO$~\cite{budivsic2012applied,mezic2015applications}. If $\calF$ is an invariant subspace, then $r$ can be zero~\cite{brunton2016koopman}.

To recover $w_{t}$ from $\Phi\left( w_{t} \right)$, we often use $w_{t}$ as part of observables such that $w_{t} = C_k \Phi\left( w_{t} \right)$, where $C_k = [\mathbf{I}_{n_x}, \; \mathbf{0}]$.

In the case where $h$ also depends on $z_t$, \cref{eq:koopman_auto} can be extended as~\cite{abraham2019active,jia2023evolver}
\begin{equation}
     \Phi\left( w_{t+1} \right) = A_k \Phi\left( w_{t} \right) + B_k \Psi\left( w_{t}, z_{t+1} \right) + r.
     \label{eq:koopman_nonauto}
\end{equation}

We assume the following for the residual dynamics $h$.

\begin{assumption}[Linear Representation of $h$ by Koopman Operator]\label{assumption:koopman}
There exist nonlinear functions $\Phi: \mathbb{R}^{d_x} {\rightarrow} \mathbb{R}^{d_\Phi} $ and $\Psi: \mathbb{R}^{d_x} \times \mathbb{R}^{d_z} \rightarrow \mathbb{R}^{d_\Psi}$ such that 
\begin{equation}
    \Phi\left( w_{t} \right) = A_{k} \Phi\left( w_{t-1} \right) + B_{k} \Psi\left( w_{t-1}, z_{t} \right),
\end{equation}
where $A_{k}: \mathbb{R}^{d_\Phi} \times \mathbb{R}^{d_\Phi}$, $B_{k}: \mathbb{R}^{d_\Phi} \times \mathbb{R}^{d_\Psi}$, and $\Phi\left( \cdot \right)$ are locally Lipschitz, and $\Psi\left( \cdot, \cdot \right)$ is uniformly bounded.
\end{assumption}

We require $\Psi\left( \cdot, \cdot \right)$ to be uniformly bounded as we assume that $w_{t}$ is bounded, and therefore $\Phi\left( w_{t} \right)$ is bounded, which means $\Psi\left( \cdot, \cdot \right)$ is also bounded. Examples of such function include, \eg $\sin\left(\cdot\right)$, $\cos\left(\cdot\right)$, and $\tanh\left(\cdot\right)$.

\Cref{assumption:koopman} assumes  that $r=0$ and that we can represent $h$ by a linear model given in \cref{eq:koopman_nonauto}.   

\begin{remark}[Approximation Error]
    When the approximation error $r \neq 0$, we can generalize the regret guarantee in \Cref{theorem:regret_OLMPC} using \cite[Corollary~1]{zhou2025simultaneous} such that it depends on the approximation error.
\end{remark}

\myParagraph{Control Performance Metric} We design $u_t$ to ensure a control performance that is comparable to an optimal clairvoyant (non-causal) policy that knows the residual dynamics a priori. Particularly, we consider the metric below.

\begin{definition}[Dynamic Regret]\label{def:DyReg_control}
Assume a total time horizon of operation $T$, and loss functions $c_t$, $t=1,\ldots, T$. Then, \emph{dynamic regret} is defined as
\begin{equation}
	\DReg = \sum_{t=1}^{T} c_{t}\left(x_{t}, u_{t}, w_{t}\right)-\sum_{t=1}^{T} c_{t}\left(x_{t}^{\star}, u_{t}^{\star}, w_{t}^\star\right),
	\label{eq:DyReg_control}
\end{equation}
where the cost $c_t$ explicitly dependent on the residual dynamics $w_{t}$, $u_{t}^{\star}$ is the optimal control input in hindsight, \ie the optimal (non-causal) input given a priori knowledge of the unknown $h$, and $x_{t+1}^{\star}$ is the state reached by applying the optimal control inputs $\left(u_{1}^{\star}, \; \dots, \; u_{t}^{\star}\right)$, and $w_{t}^\star$ is the residual dynamics experienced by the optimal controller.
\end{definition}

\begin{remark}[Adaptivity of $h$]\label{rem:adaptivity}
In the definition of regret in \cref{eq:DyReg_control}, $h$ adapts (possibly differently) to the state and control sequences $(x_{1},u_1), \; \dots, \;(x_{T}, u_T)$ and $(x_{1}^{\star}, u_1^\star), \; \dots, \; (x_{T}^{\star},u^\star_T)$ since $h$ is a function of the state and the control input.
    This is in contrast to previous definitions of dynamic regret, \eg \cite{hazan2022introduction,zhou2023efficient,zhou2023safecdc,zhou2023saferal} and references therein, where the optimal state $x_{t+1}^{\star}$ is reached given the same realization of $w$ as of $x_{t+1}$, \ie $x_{t+1}^{\star} = f\left(x_{t}^{\star}\right) + g\left(x_{t}^{\star}\right) u_{t}^{\star} + w_{t}$.
\end{remark}

\begin{problem}[Model Predictive Control with Online Learning of Koopman Operator]\label{prob:control}
At each $t=1,\ldots, T$, estimate the residual dynamics function ${h}$, and identify a control input $u_t$ by solving \cref{eq:mpc_ada_def}, such that $\DReg$ is sublinear.
\end{problem}

A sublinear regret means $\lim_{T\rightarrow\infty} \DReg/T \rightarrow 0$, which implies the algorithm asymptotically converges to the optimal (non-causal) controller.


\section{Algorithm}\label{sec:alg}

We present the algorithm for \Cref{prob:control} (\Cref{alg:MPC}). The algorithm is sketched in \Cref{fig_framework}.  The algorithm is composed of two interacting modules: (i) an \MPC module, and (ii) an online Koopman operator learning module.  At each $t=1,2,\ldots,$ the \MPC module uses the estimated ${h}$ from the Koopman operator learning module to calculate the control input $u_t$. Given the current control input $u_t$ and the observed new state $x_{t+1}$, the Koopman operator learning module updates the estimate ${h}$.  To this end, it employs online least-squares estimation via online gradient descent, where $h$ is parameterized as a linear system by Koopman operator. 

To present the algorithm, we first next provide background information on online gradient descent for estimation.

\subsection{Online Least-Squares Estimation}\label{subsec:OLS}
Given a data point $\left(w_{t-1}, \; z_t, \; w_{t} \right)$ observed at time $t$, we employ an online least-squares algorithm that updates the parameters $\hat{\alpha}_t \triangleq \left[ \hat{A}_{k,t}, \;  \hat{B}_{k,t}\right]$ to minimize the approximation error $l_t = \| \Phi\left( w_{t} \right) - \Phi\left( \hat{w}_{t}  \right)\|_2^2$, where $ \Phi\left( \hat{w}_{t} \right) = \hat{A}_{k,t} \Phi\left( w_{t-1} \right) + \hat{B}_{k,t} \Psi\left( w_{t-1}, z_{t} \right)$. 
Specifically, the algorithm uses the online gradient descent algorithm~(\OGD)~\cite{hazan2016introduction}. At each $t = 1, \dots, T$, it makes the steps:
\begin{itemize}
    \item Given $\left(w_{t-1}, \; z_t, \; w_{t} \right)$, formulate the estimation loss function (approximation error):
            \begin{equation*}
                l_t\left(\hat{\alpha}_t\right) \triangleq \left\| \Phi\left( w_{t} \right) -  \hat{A}_{k,t} \Phi\left( w_{t-1} \right) - \hat{B}_{k,t} \Psi\left( w_{t-1}, z_{t} \right) \right\|^2.
            \end{equation*}
    \item Calculate the gradient of $l_t\left(\hat{\alpha}_t\right)$ with respect to $\hat{\alpha}_t$: 
            \begin{equation*}
                \nabla_t \triangleq \nabla_{\hat{\alpha}_t} l_t\left(\hat{\alpha}_t\right).
            \end{equation*}
    \item Update using gradient descent with learning rate $\eta$:
            \begin{equation*}
                \hat{\alpha}_{t+1}^\prime= \hat{\alpha}_t- \eta \nabla_t.
            \end{equation*}
    \item Project $\hat{\alpha}_{t+1}^\prime$ onto a convex and compact domain set $\calD$:
            \begin{equation*}
                \hat{\alpha}_{t+1} = \Pi_{\calD}(\hat{\alpha}_{t+1}^\prime) \triangleq \underset{\alpha \in \calD}{\operatorname{\textit{argmin}}}\; \| \alpha - \hat{\alpha}_{t+1}^\prime \|^2_2.
            \end{equation*}
\end{itemize}

The above online least-squares estimation enjoys an $\calO\left(\sqrt{T}\right)$ regret bound, per the regret bound of \OGD~\cite{hazan2016introduction}.

\begin{proposition}[Regret Bound of Online Least-Squares Estimation~\cite{hazan2016introduction}]\label{theorem:OGD}
    Assume $\eta=\calO\left({1}/{\sqrt{T}}\right)$.  Then,
    \begin{equation}
       \SReg\triangleq \sum_{t=1}^{T} l_t \left(\alpha_t\right) - \sum_{t=1}^{T} l_t \left(\alpha^{\star}\right)  \leq \calO\left(\sqrt{T}\right),
    \end{equation}
    where $\alpha^{\star} \triangleq \underset{\alpha \in \calD}{\operatorname{\textit{argmin}}}\;\sum_{t=1}^{T} l_t \left(\alpha\right)$ is the optimal parameter that achieves lowest cumulative loss in hindsight.
\end{proposition}

The online least-squares estimation algorithm thus asymptotically achieves the same estimation error 
as the optimal parameter $\alpha^{\star}$ since $\lim_{T\rightarrow\infty} \;\SReg/T = 0$.

\subsection{Algorithm for \Cref{prob:control}}\label{subsec:MPC}

\setlength{\textfloatsep}{-0.1mm}
\begin{algorithm}[t]
\small
	\caption{Simultaneous Koopman Learning and Model Predictive Control (\KMMPC).}
	\begin{algorithmic}[1]
		\REQUIRE Koopman observable functions $\Phi$ and $\Psi$;  domain set $\calD$;  gradient descent learning rate $\eta$.
		\ENSURE At each time step $t=1,\ldots,T$, control input $u_{t}$.
		\medskip
            \STATE Initialize $x_1$, $\hat{\alpha}_{1} \in \calD$; 
		\FOR {each time step $t = 1, \dots, T$}
		\STATE Apply control input $u_t$ by solving \cref{eq:mpc_ada_def} with ${h}\left(\cdot, \cdot; \hat{\alpha}_{t}\right) \triangleq  C_k \left( \hat{A}_{k,t} \Phi\left( \cdot \right) + \hat{B}_{k,t} \Psi\left( \cdot, \cdot \right) \right)$;
            \STATE Observe state $x_{t+1}$, and calculate residuals via $w_{t} = x_{t+1} - f(x_{t}) - g(x_{t}) u_{t}$;
            \STATE Formulate estimation loss $l_t\left(\hat{\alpha}_t\right) \triangleq \left\| \Phi\left( w_{t} \right) -  \hat{A}_{k,t} \Phi\left( w_{t-1} \right) - \hat{B}_{k,t} \Psi\left( w_{t-1}, z_{t} \right) \right\|^2$;
            \STATE Calculate gradient $\nabla_t \triangleq \nabla_{\hat{\alpha}_t} l_t\left(\hat{\alpha}_t\right)$;
            \STATE Update $\hat{\alpha}_{t+1}^\prime= \hat{\alpha}_t- \eta \nabla_t$;
            \STATE Project  $\hat{\alpha}_{t+1}^\prime$ onto $\calD$, \ie $\hat{\alpha}_{t+1} = \Pi_{\calD}(\hat{\alpha}_{t+1}^\prime)$;
            \ENDFOR
	\end{algorithmic}\label{alg:MPC}
\end{algorithm}

We describe the algorithm for \Cref{prob:control}. The pseudo-code is in \Cref{alg:MPC}.  The algorithm is composed of three steps: initialization, control, and estimation. The control and estimation steps are interacting and influence each other at each time step (Fig.~\ref{fig_framework}):

\begin{itemize}
    \item \textit{Initialization step:} \Cref{alg:MPC} first initializes the system state $x_1$ and parameter $\hat{\alpha}_1 \in \calD$~(line 1).

    \item \textit{Control steps:} Then, at each~$t$, given the current estimate ${h}\left(\cdot, \cdot; \hat{\alpha}_{t}\right) \triangleq  C_k \left( \hat{A}_{k,t} \Phi\left( \cdot \right) + \hat{B}_{k,t} \Psi\left( \cdot, \cdot \right) \right)$, \Cref{alg:MPC} applies the control inputs $u_t$ by solving \cref{eq:mpc_ada_def}~(line 3).

    \item \textit{Estimation steps:}
    The system then evolves to state $x_{t+1}$, and, $w_t$ is calculated upon observing $x_{t+1}$~(line 4).
    Afterwards, the algorithm formulates the loss $l_t\left(\hat{\alpha}_t\right) \triangleq \left\| \Phi\left( w_{t} \right) -  \hat{A}_{k,t} \Phi\left( w_{t-1} \right) - \hat{B}_{k,t} \Psi\left( w_{t-1}, z_{t} \right) \right\|^2$~(line 5), and calculates the gradient $\nabla_t \triangleq \nabla_{\hat{\alpha}_t} l_t\left(\hat{\alpha}_t\right)$~(line 6). 
    \Cref{alg:MPC} then updates the parameter $\hat{\alpha}_t$ to $\hat{\alpha}_{t+1}^\prime$~(line 8) and projects $\hat{\alpha}_{t+1}^\prime$ back to the domain $\calD$~(line 9).
\end{itemize}

\begin{remark}[Combination with Offline Learned Koopman Operator]\label{remark:offline}
    The proposed online learning method can be combined with offline learning of Koopman operator. For example, we can learn a linear representation of the nominal dynamics as $\bar{\Phi}\left( x_{t+1} \right) = \bar{A}_k \bar{\Phi}\left( x_{t} \right) + \bar{B}_k u_k$~\cite{korda2018linear} and use \OGD to learn online the residual dynamics in the lift space of the form $\Delta \bar{A}_k \bar{\Phi}\left( x_{t} \right) + \Delta\bar{B}_k u_k$. 
    To obtain \Cref{theorem:regret_OLMPC}, we require \Cref{assumption:stability}, \Cref{assumption:lipschitz}, and uniform boundedness of $x_t$. We replace \Cref{assumption:koopman} by the assumption of uniform boundedness of $x_t$ since $\bar{\Phi}\left( \cdot \right)$ typically is not an uniformly bounded function and can violate the assumption of bounded $w_t$, which may cause the loss and gradient (lines 5-6 in \Cref{alg:MPC}) to be unbounded.
\end{remark}

\section{No-Regret Guarantee}\label{sec:Reg}

We present the sublinear regret bound of \Cref{alg:MPC}.

\begin{theorem}[No-Regret]\label{theorem:regret_OLMPC}
Assume \Cref{alg:MPC}'s learning rate is $\eta=\calO\left({1}/{\sqrt{T}}\right)$.  Then, \Cref{alg:MPC} achieves  
\begin{equation}
    \DReg \leq \calO\left(T^{\frac{3}{4}}\right).
    \label{eq:theorem_regret_OLMPC}
\end{equation}
\end{theorem}


The proof follows the steps in the proof of \cite[Theorem~1]{zhou2025simultaneous}

\Cref{theorem:regret_OLMPC} serves as a finite-time performance guarantee as well as implies that \Cref{alg:MPC} converges to the optimal (non-causal) control policy since $\lim_{T\rightarrow\infty}\DReg / T \rightarrow 0$.

\section{Numerical Evaluations}\label{sec:exp-sim}

We evaluate \Cref{alg:MPC} in simulated scenarios of control under uncertainty, where the controller aims to track a reference setpoint despite unknown residual dynamics. 
Specifically, we consider a cart-pole aiming to stabilize around a setpoint despite inaccurate model parameters, \ie inaccurate cart mass, pole mass, and pole length.

\begin{figure}[!]
    \centering
    \subfigure[Average stabilization error.]{\includegraphics[width=0.4\textwidth]{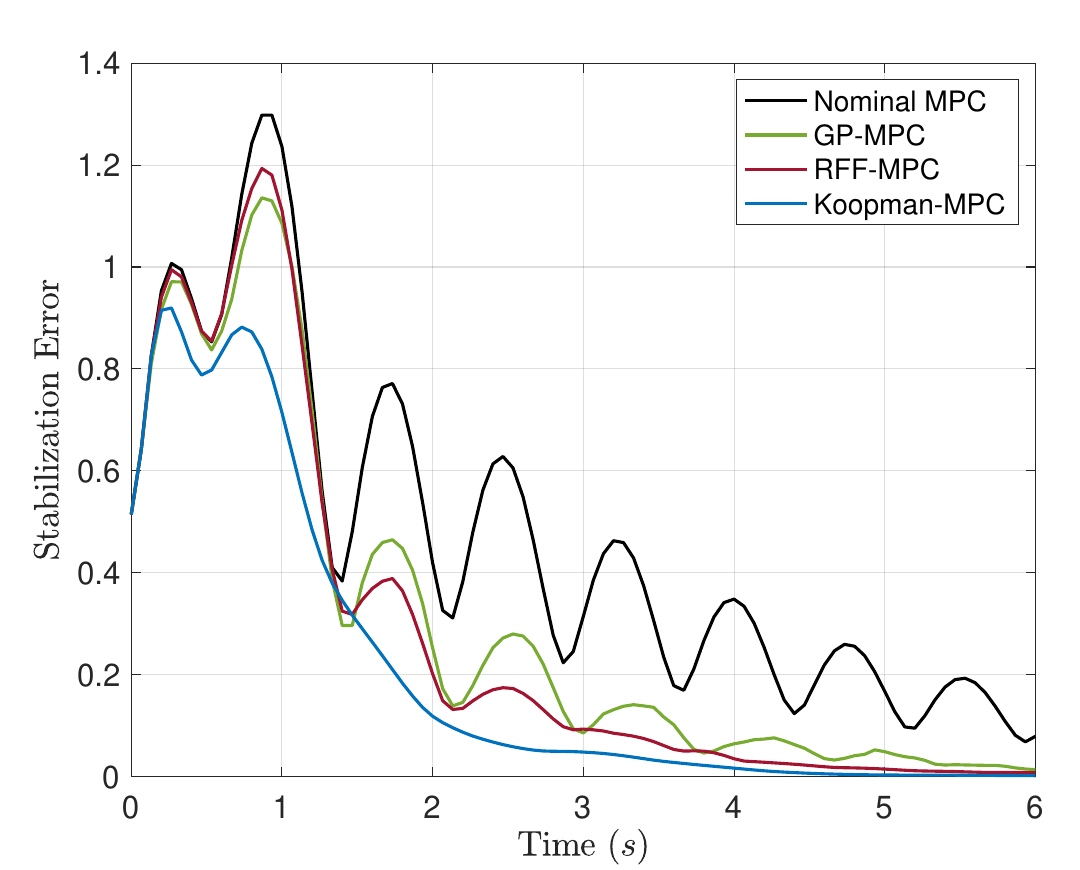}}\label{fig_cartpole_error}
    \subfigure[Sample trajectory.]{\includegraphics[width=0.4\textwidth]{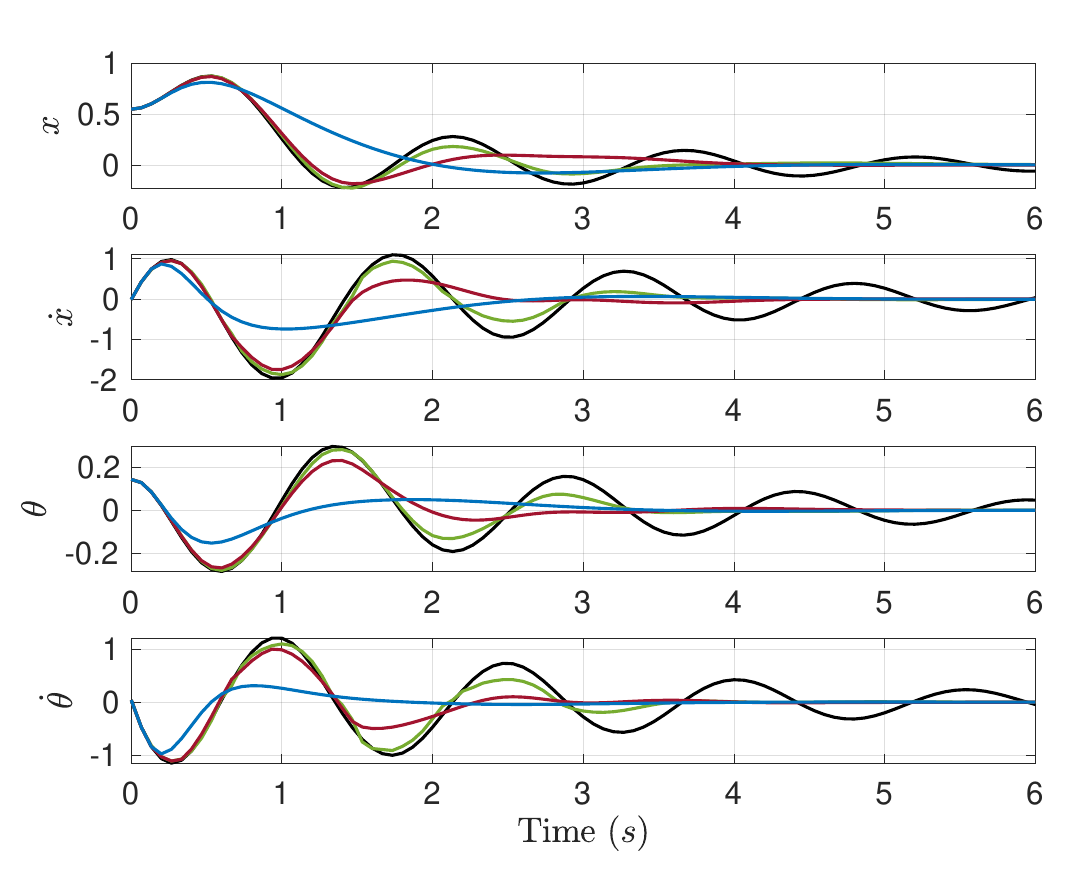}\label{fig_cartpole_traj}} 
    \subfigure[Prediction error.]{\includegraphics[width=0.4\textwidth]{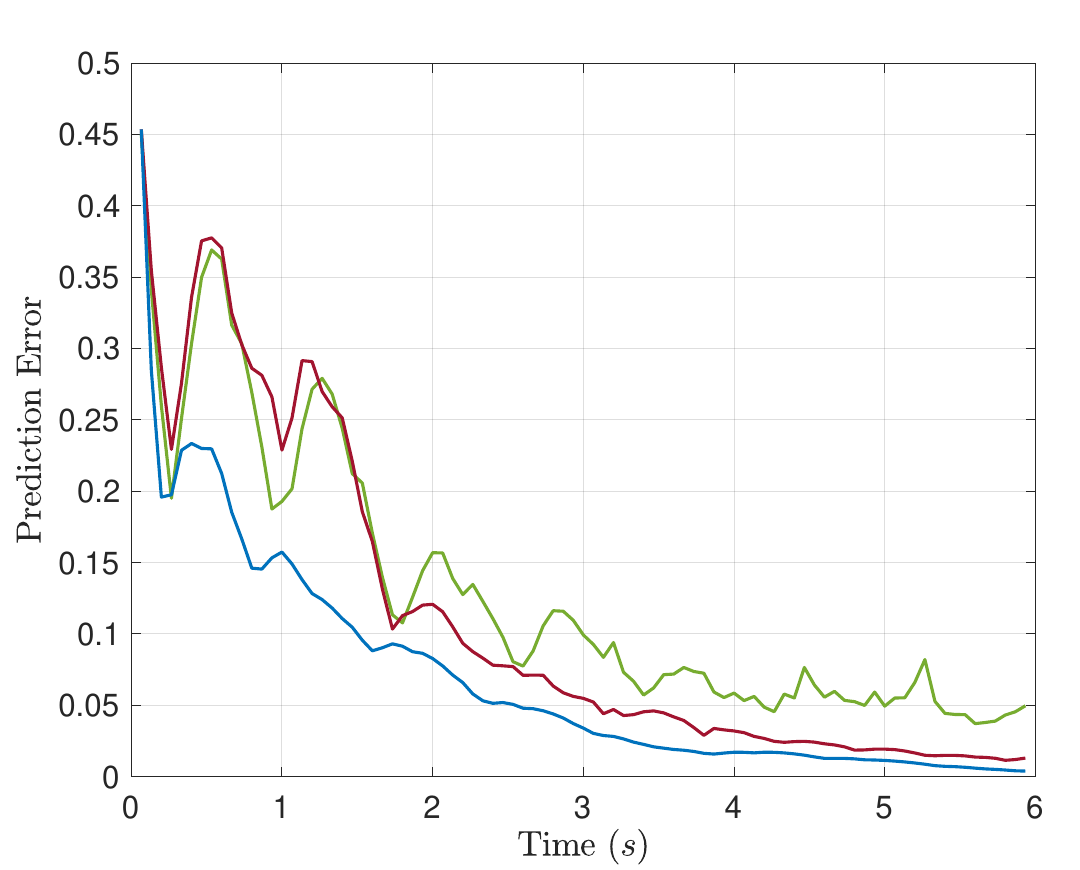}\label{fig_cartpole_pred}}
    \caption{\textbf{Simulation Results of the Cart-Pole Stabilization Experiment under $25\%$ Inaccurate Model Parameters.} (a) Average stabilization error over $20$ runs with random initialization. (b) Sample trajectory. The results demonstrate that \Cref{alg:MPC}~(\KMMPC) achieves the fastest stabilization of the system among all tested algorithms. (c) Estimation error of the residual dynamics. The results demonstrate that the quick convergence of online learning of the Koopman operator with appropriately chosen observables.}
    \label{fig_cartpole_stabilization}
\end{figure}

\begin{figure}[t]
    \centering
    \includegraphics[width=0.49\textwidth]{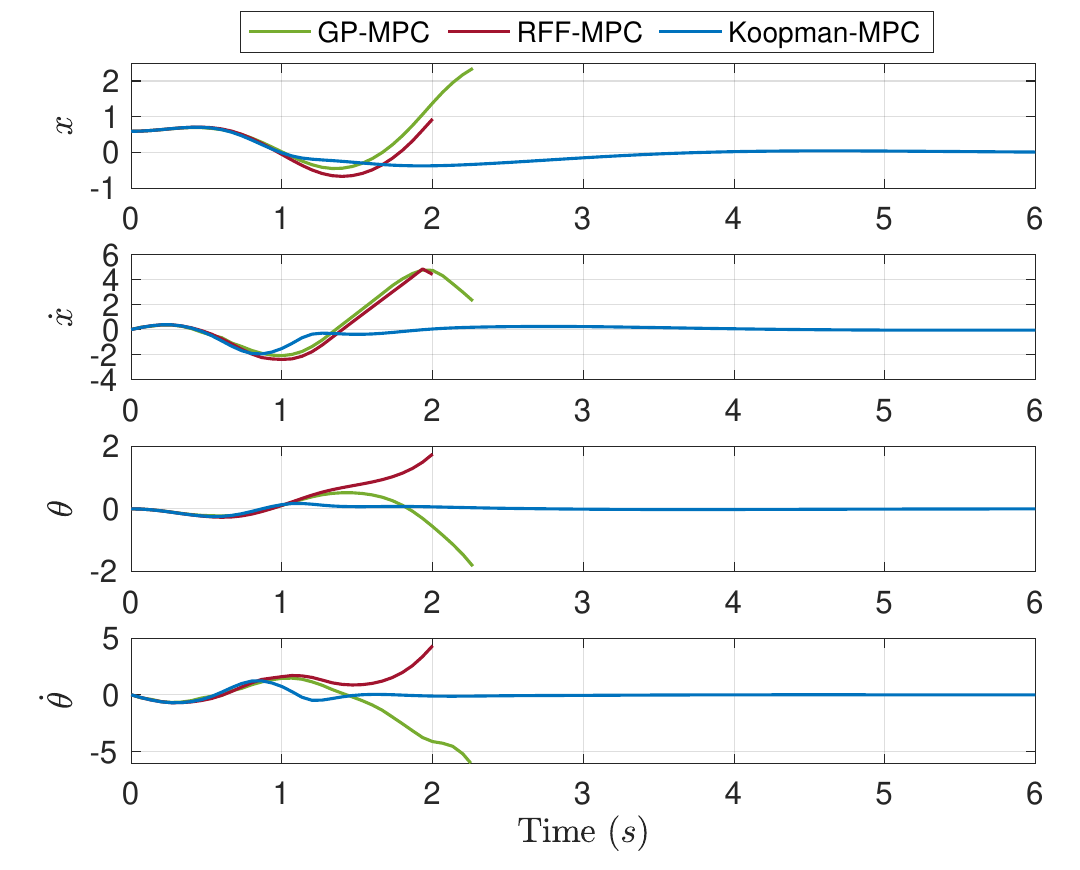}
    \caption{\textbf{Sample Trajectory of the Cart-Pole Stabilization Experiment under $45\%$ Inaccurate Model Parameters.} \Cref{alg:MPC}~(\KMMPC) successfully achieves stabilization of the system despite $45\%$ inaccuracy of the nominal model, while \GPMPC and \RFFMPC fail.}
    \label{fig_cartpole_055}
    \vspace{5mm}
\end{figure}

\myParagraph{Simulation Setup} We consider a cart-pole system, where a cart of mass $m_c$ connects via a prismatic joint to a $1D$
track, while a pole of mass $m_p$ and length $2l$ is hinged to the cart.
The state vector $\myx$ includes the horizontal position of the cart $x$, the velocity of the cart $\dot{x}$,  the angle of the pole with respect to vertical $\theta$, and the angular velocity of the pole $\dot{\theta}$.
The control input is the force $F$ applied to the center of mass of the cart. 
The goal of the cart-pole is to stabilize at $(x, \dot{x}, \theta, \dot{\theta}) = (0,0,0,0)$. The cart-pole dynamics are:
\begin{equation}
    \begin{aligned}
        \ddot{x} &= \frac{m_p l \left(\dot{\theta}^2 \sin\theta - \ddot{\theta} \cos\theta \right) + F}{m_c + m_p}, \\
        \ddot{\theta} &= \frac{g \sin\theta + \cos\theta \left( \frac{- m_p l \dot{\theta}^2 \sin\theta - F}{m_c + m_p} \right) }{l \left( \frac{4}{3} - \frac{m_p \cos^2\theta}{m_c + m_p} \right)} ,
    \end{aligned}
\end{equation}
where $g$ is the acceleration of gravity.

To control the system, we will employ \MPC at $15Hz$ with a look-ahead horizon $N=20$. We use quadratic cost functions with $Q = \diag{[5.0,\; 0.1,\; 5.0,\; 0.1]}$ and $R=0.1$. We use the fourth-order Runge-Kutta method for discretizing the above dynamics. 
The true system parameters are $m_c=1.0$, $m_p=0.1$, and $l=0.5$, but the parameters for the nominal dynamics are scaled to $75\%$ and $55\%$ of the said true values, which corresponding to $25\%$ and $45\%$ inaccuracy.


We use $\Phi\left(w_{t}\right) = w_{t}$, $\Psi\left(w_{t-1}, z_{t}\right) = \left[ \tanh{x_t}, \; \tanh{\dot{x}_t}, \; \tanh{\theta_t}, \; \tanh{\dot{\theta}_t}, \; u_t, \; \tanh{\theta_t} \tanh{\dot{\theta}_t}, \right. $ $\left. (\tanh{\dot{\theta}_t})^2, \; u_t \tanh{\theta}_t , \; u_t \tanh{\dot{\theta}_t}  \right]^\top$, $\eta=0.01$, and initialize $\hat{\alpha}$ as zero.
We simulate the setting for $6$ seconds, performing the simulation $20$ times with random initialization of the state sampled uniformly from $x \in [-1,\; 1]$, $\dot{x} \in [-0.1,\; 0.1]$, $\theta \in [-0.2,\; 0.2]$, $\dot{\theta} \in [-0.1,\; 0.1]$.

We use the physics-based simulation environment from~\cite{yuan2022safe}
in PyBullet~\cite{coumans2016pybullet}.

\myParagraph{Compared Algorithms} 
We compare \Cref{alg:MPC}~(\KMMPC) with an \MPC  that uses the nominal system parameters (\NMPC), the Gaussian process \MPC (\GPMPC)~\cite{hewing2019cautious}, and \MPC with random Fourier Features (\RFFMPC)~\cite{zhou2025simultaneous}.
The \NMPC uses the nominal dynamics to select control input by solving \cref{eq:mpc_def}.
The \GPMPC learns $\hat{h}\left(\cdot\right)$ with a sparse Gaussian process~(\GP)~\cite{quinonero2005unifying} whose data points are collected online, \ie~\GP fixes its hyperparameters and collects data points $\left(w_{t-1}, \; z_t, \; w_{t} \right)$ online.
The \RFFMPC learns $\hat{h}\left(\cdot\right)$ with random Fourier features~\cite{rahimi2007random} also with data points $\left(w_{t-1}, \; z_t, \; w_{t} \right)$ collected online.

\myParagraph{Performance Metric} 
We evaluate the performance of \NMPC, \GPMPC, \RFFMPC, and \Cref{alg:MPC} in terms of their stabilization error $\|\myx_t\|^2$.

\myParagraph{Results} The results are given in \Cref{fig_cartpole_stabilization} and \Cref{fig_cartpole_055}. 

In the case of $25\%$ inaccurate model parameters (\Cref{fig_cartpole_stabilization}), we observe that \Cref{alg:MPC} achieves stabilization the fastest. 
\RFFMPC comes second, possibly because it estimates residual dynamics with randomly sampled features while our method uses suitable observables which enables better learning of residual dynamics~(\Cref{fig_cartpole_pred}). 
\GPMPC is able to stabilize the system at the end but it incurs a larger deviation during the transition period from the stabilization goal $(0,0,0,0)$ than \Cref{alg:MPC}.

In the case of $45\%$ inaccurate model parameters (\Cref{fig_cartpole_055}), \Cref{alg:MPC} successfully stabilizes the system in all $20$ runs, while \GPMPC and \RFFMPC fail. The result demonstrates robustness of \Cref{alg:MPC} under such extreme disturbances.

\section{Conclusion} \label{sec:con}

\myParagraph{Summary}
We provided \Cref{alg:MPC} for the problem of \textit{Model Predictive Control with Online Learning of Koopman Operator}~(\Cref{prob:control}). 
\Cref{alg:MPC} guarantees no-regret against an optimal clairvoyant policy that knows the residual dynamics $h$ a priori. (\Cref{theorem:regret_OLMPC}). 
The algorithm uses Koopman operator to approximate the residual dynamics. Then, it employs model predictive control based on the current learned model of $h$.
The model of the unknown dynamics is updated online in a self-supervised manner using least squares based on the data collected while controlling the system.
We validate \Cref{alg:MPC} in physics-based PyBullet simulations of a cart-pole aiming to maintain the pole upright despite inaccurate model parameters~(\Cref{sec:exp-sim}).
We demonstrate that our method achieves better tracking performance than the state-of-the-art methods \GPMPC~\cite{hewing2019cautious} and \RFFMPC~\cite{zhou2025simultaneous}. 

\myParagraph{Future Work}
The Koopman observable functions in the paper are manually selected, and may not generalize to settings with different unknown dynamics or disturbances. We expect this can be resolved by using meta-learning (with neural networks) to automate the discovery of Koopman observable functions that can generalize to different $h\left(\cdot\right)$. 



\bibliographystyle{IEEEtran}
\bibliography{References}



\end{document}